\documentclass[prd,onecolumn,notitlepage,eqsecnum,nofootinbib,floatfix]{revtex4-1}
\usepackage{amssymb}
\usepackage{amsmath}
\usepackage{amsfonts} 
\allowdisplaybreaks
\begin{document}
\title{EZ gauge is singular at the event horizon} 
\author{Eamonn Corrigan and Eric Poisson}  
\affiliation{Department of Physics, University of Guelph, Guelph,
  Ontario, N1G 2W1, Canada} 
\date{May 10, 2018} 
\begin{abstract} 
We prove that the EZ gauge of black-hole perturbation theory,
introduced by the late Steve Detweiler in his exploration of the
physical consequences of the gravitational self-force, is necessarily
singular at the event horizon. 
\end{abstract} 
\maketitle

\section{Introduction and summary} 
\label{sec:intro} 

Black-hole perturbation theory, understood here in the specific guise
of metric perturbations of the Schwarzschild spacetime, is a mature
framework that has found many fruitful applications. The
theory originated in the works of Regge and Wheeler
\cite{regge-wheeler:57}, Vishveshwara \cite{vishveshwara:70}, and
Zerilli \cite{zerilli:70}, and it was formalized in gauge-invariant
formalisms by Moncrief \cite{moncrief:74}, Gerlach and Sengupta
\cite{gerlach-sengupta:79, gerlach-sengupta:80}, Gundlach and
Martin-Garcia \cite{gundlach-martingarcia:00,
  gundlach-martingarcia:01}, Sarbach and Tiglio
\cite{sarbach-tiglio:01}, Clarkson and Barrett 
\cite{clarkson-barrett:03}, Nagar and collaborators
\cite{nagar-etal:04, nagar-rezzolla:05}, and Martel and Poisson
\cite{martel-poisson:05}. The theory was applied to a plethora of
phenomena, including the quasinormal modes of vibration of a black
hole \cite{kokkotas-schmidt:99}, the gravitational waves produced by a
point particle in orbit around a black hole \cite{sasaki-tagoshi:03,
  martel:04}, the self-force acting on a small body inspiralling
toward a black hole \cite{barack:09, warburton-etal:12}, the collision
of two black holes in a close-limit approximation
\cite{gleiser-etal:00}, and the tidal deformation of black holes
\cite{poisson:04d, poisson:05, poisson-vlasov:10}. 

The equations of black-hole perturbation theory can be integrated by
making use of gauge-invariant master functions, or by selecting
a specific gauge to eliminate the redundant coordinate freedom. The
most popular choice has been the Regge-Wheeler gauge, introduced in 
Ref.~\cite{regge-wheeler:57}, which possesses the virtues of being
algebraically simple and unique (except for the monopole and dipole
pieces of the perturbation). Another choice, strongly embraced by one
of the authors in his exploration of the tidal deformation of black
holes, is the light-cone gauge of Preston and Poisson
\cite{preston-poisson:06b}; this gauge assigns a compelling 
geometrical meaning to the coordinates of the perturbed spacetime, but
it is not unique. Another choice of gauge was recently contributed to
the vast literature on black-hole perturbation theory. It is known as
the EZ (easy) gauge \cite{thompson-chen-whiting:17}, and it was
devised (but never published) by the late Steve Detweiler during his
exploration of the physical consequences of the gravitational
self-force \cite{detweiler:08}.  

The choice of gauge is entirely a matter of taste and convenience; 
metric perturbations related by a gauge transformation are physically 
equivalent. The choice, however, should be informed by the known
properties of the gauge, and our purpose in this paper is to point
out that {\it in the EZ gauge, the metric perturbation is necessarily
  singular on the black-hole horizon}. The statement is true for static
and time-dependent situations, and is true regardless of the source of
the perturbation. This fact does not seem to have been noticed
before. The singularity of the EZ gauge at the event horizon does not
imply that the gauge is necessarily ``bad'' or that its use should be 
discouraged. After all, one can be perfectly comfortable working with
the Schwarzschild metric in the standard $(t,r,\theta,\phi)$
coordinates, in spite of the metric singularity at $r=2M$; one knows
to be careful when investigating processes that occur at or near 
the horizon. The same lesson applies to the EZ gauge: some care is
required if one wishes to adopt it. For example, the EZ gauge can 
fruitfully be employed to calculate the gravitational waves emitted by
a particle in orbit around a black hole, since the metric perturbation
in this gauge can readily be related to the gauge-invariant master
functions, which have a close relation to the two gravitational-wave 
polarizations. Such an application was considered in Sec.~10 of
Ref.~\cite{thompson-chen-whiting:17}. 

To establish the singular property of the EZ gauge, we begin with the
unperturbed Schwarzschild metric presented in $(v,r,\theta,\phi)$
coordinates, with $v$ denoting the standard advanced-time
coordinate. The metric is known to be regular at $r=2M$ in these
coordinates. We next introduce a metric perturbation formulated in the 
light-cone gauge \cite{preston-poisson:06b}, which preserves the
geometrical meaning of the spacetime coordinates. This property makes
it self-evident that in the light-cone gauge, the components of a
physically regular metric perturbation will be nonsingular at
$r=2M$. This perturbation is then transformed to the Regge-Wheeler
gauge \cite{regge-wheeler:57}, and shown again to have nonsingular
components at the horizon. Finally, the perturbation is transformed to
the EZ gauge, and shown to have components that diverge 
at $r=2M$. A physically regular metric perturbation therefore appears
to be singular in the EZ gauge. (We note that the proof could avoid
the middle step of transforming to the Regge-Wheeler gauge. But we
find it useful to establish the regularity of this gauge as a
byproduct of the proof.) 

We begin in Sec.~\ref{sec:perturbation} with a brief summary of the
essential points of black-hole perturbation theory. In
Sec.~\ref{sec:LCRW} we consider a perturbation presented in the
light-cone gauge and construct the transformation that takes it to the
Regge-Wheeler gauge; we observe that the perturbation stays
regular at $r=2M$ after the transformation. In Sec.~\ref{sec:RWEZ} we
construct the transformation to the EZ gauge, and show that the
perturbation now diverges at the event horizon. Finally, as
a concrete illustration of these properties, we display in
Sec.~\ref{sec:tidal} the perturbations that correspond to a black hole 
deformed by a quadrupolar tidal field.    

\section{Metric perturbation and gauge transformations} 
\label{sec:perturbation} 

We rely on the summary of black-hole perturbation theory provided by
Martel and Poisson \cite{martel-poisson:05}. The unperturbed metric
$g_{\alpha\beta}$ is given by the Schwarzschild solution in
$(v,r,\theta,\phi)$ coordinates, 
\begin{equation} 
g_{\alpha\beta}\, dx^\alpha dx^\beta 
= g_{ab}\, dx^a dx^b + r^2 \Omega_{AB}\, d\theta^A d\theta^B, 
\end{equation} 
where $x^a := (v,r)$, $\theta^A := (\theta,\phi)$, $\Omega_{AB} :=
\mbox{diag}[1,\sin^2\theta]$ is the metric on a unit 2-sphere, and 
\begin{equation}  
g_{ab}\, dx^a dx^b = -f\, dv^2 + 2dvdr
\end{equation} 
with $f := 1-2M/r$; $M$ denotes the mass of the black hole. The
perturbed metric is $g_{\alpha\beta} + p_{\alpha\beta}$, with
$p_{\alpha\beta}$ representing the perturbation. 

The perturbation is decomposed in spherical harmonics according to 
\begin{subequations} 
\begin{align} 
p_{ab} &= \sum_{\ell m} h_{ab}^{\ell m} Y^{\ell m}, \\ 
p_{aB} &= \sum_{\ell m} j_a^{\ell m} Y_A^{\ell m} 
+ \sum_{\ell m} h_a^{\ell m} X_A^{\ell m}, \\ 
p_{AB} &= r^2 \sum_{\ell m} \bigl( K^{\ell m} \Omega_{AB} Y^{\ell m} 
+ G^{\ell m} Y^{\ell m}_{AB} \bigr) 
+ \sum_{\ell m} h_2^{\ell m} X_{AB}^{\ell m},  
\end{align} 
\end{subequations} 
where $Y^{\ell m}(\theta^A)$ are the usual scalar harmonics, 
$Y^{\ell m}_A$ and $Y^{\ell m}_{AB}$ are even-parity harmonics, and  
$X^{\ell m}_A$ and $X^{\ell m}_{AB}$ are odd-parity harmonics;
definitions are provided in Sec.~III of
Ref.~\cite{martel-poisson:05}. The perturbation fields 
$h_{ab}^{\ell m}$, $j_a^{\ell m}$, $K^{\ell m}$, and $G^{\ell m}$, all
functions of $x^a$, make up the even-parity sector of the
perturbation. The fields $h_a^{\ell m}$ and $h_2^{\ell m}$, also
functions of $x^a$, make up the odd-parity sector of the
perturbation. In the decomposition we assume that the sums over $\ell$
begin with $\ell = 2$. We exclude the monopole and dipole pieces of
the perturbation because both the Regge-Wheeler and EZ gauges are not
defined for them; alternative gauge choices must be made.   

A gauge transformation is generated by a vector field $\Xi_\alpha$
that can be decomposed as 
\begin{subequations} 
\begin{align} 
\Xi_a &= \sum_{\ell m} \xi_a^{\ell m} Y^{\ell m}, \\ 
\Xi_A &= \sum_{\ell m} \xi_{\rm even}^{\ell m} Y_A^{\ell m} 
+ \sum_{\ell m} \xi_{\rm odd}^{\ell m} X_A^{\ell m}. 
\end{align} 
\end{subequations} 
The transformation produces the changes
\begin{subequations} 
\label{gauge_even} 
\begin{align} 
h_{vv} &\to h_{vv} 
- 2 \partial_v \xi_v 
+ \frac{2M}{r^2} \xi_v + \frac{2Mf}{r^2} \xi_r, \\ 
h_{vr} &\to h_{vr} 
- \partial_r \xi_v 
- \partial_v \xi_r - \frac{2M}{r^2} \xi_r, \\
h_{rr} &\to h_{rr} - 2 \partial_r \xi_r, \\
j_v &\to j_v 
- \partial_v \xi_{\rm even} - \xi_v, \\ 
j_r &\to j_r 
- \partial_r \xi_{\rm even} - \xi_r 
+ \frac{2}{r} \xi_{\rm even}, \\
K &\to K 
-\frac{2f}{r} \xi_r - \frac{2}{r} \xi_v 
+ \frac{\ell(\ell+1)}{r^2} \xi_{\rm even}, \\ 
G &\to G - \frac{2}{r^2} \xi_{\rm even}
\end{align} 
\end{subequations} 
in the even-parity sector, and 
\begin{subequations} 
\label{gauge_odd} 
\begin{align} 
h_v &\to h_v 
- \partial_v \xi_{\rm odd}, \\
h_r &\to h_r 
-  \partial_r \xi_{\rm odd}  
+ \frac{2}{r} \xi_{\rm odd}, \\ 
h_2 &\to h_2 - 2\xi_{\rm odd} 
\end{align}
\end{subequations} 
in the odd-parity sector. To unclutter the notation we omit
the $\ell m$ labels on the perturbation and gauge fields. 

\section{From light-cone gauge to Regge-Wheeler gauge} 
\label{sec:LCRW} 

The light-cone gauge of black-hole perturbation theory was introduced
by Preston and Poisson \cite{preston-poisson:06b}. Its formulation
begins with the observation that the coordinates $(v,r,\theta,\phi)$
of the background Schwarzschild spacetime possess a clear geometrical
meaning. The advanced-time coordinate $v$ is constant on light cones
that converge toward the future singularity at $r=0$, $-r$ is an
affine-parameter distance on each null generator of these light cones,
and $\theta^A = (\theta,\phi)$ is constant on these generators. It is
this compelling geometrical meaning that ensures that in these
coordinates, the Schwarzschild metric is regular at the event
horizon. The argument is simply that since the light cones behave
smoothly as they cross $r=2M$, and since the coordinates are tied 
to these light cones, the metric will be regular when presented in
these coordinates.  

The light-cone gauge places conditions on $p_{\alpha\beta}$ that
ensure that the geometrical meaning of the coordinates is preserved in
the perturbed spacetime. In this way, $v$ continues to label
converging light cones (now perturbed), $-r$ continues to be an affine
parameter on the null generators, and $\theta^A$ continues to be
constant on each generator. The preserved geometrical meaning of the
coordinates guarantees that in the light-cone gauge, the components of
$p_{\alpha\beta}$ in $(v,r,\theta,\phi)$ coordinates will be regular
at $r=2M$. The argument is the same as for the background
spacetime, and we therefore take it as self-evident that a
perturbation presented in the light-cone gauge will be regular at the
black-hole horizon. 

The light-cone gauge conditions \cite{preston-poisson:06b} are $p_{vr}
= p_{rr} = p_{rA} = 0$, so that 
\begin{equation} 
h^{\rm LC}_{vr} = h^{\rm LC}_{rr} = j^{\rm LC}_r = 0, \qquad 
h^{\rm LC}_v = 0. 
\end{equation} 
The Regge-Wheeler gauge conditions \cite{regge-wheeler:57} are 
\begin{equation} 
j^{\rm RW}_v = j^{\rm RW}_r = G^{\rm RW} = 0, \qquad 
h^{\rm RW}_2 = 0. 
\end{equation} 
When we incorporate these in Eqs.~(\ref{gauge_even}), we find that the
transformation from light-cone gauge to Regge-Wheeler gauge is
achieved with   
\begin{subequations} 
\begin{align} 
\xi_v^{\rm LC \to RW} &= j_v^{\rm LC} 
- \frac{1}{2} r^2 \partial_v G^{\rm LC}, \\ 
\xi_r^{\rm LC \to RW} &= 
-\frac{1}{2} r^2 \partial_r G^{\rm LC}, \\ 
\xi_{\rm even}^{\rm LC \to RW} &= 
\frac{1}{2} r^2 G^{\rm LC} 
\end{align} 
\end{subequations} 
in the even-parity sector. In the odd-parity sector we have 
\begin{equation} 
\xi_{\rm odd}^{\rm LC \to RW} 
= \frac{1}{2} h_2^{\rm LC}. 
\end{equation} 
We remark that the gauge vector is uniquely determined (the
Regge-Wheeler gauge is unique), and that the operators acting on the
light-cone perturbation fields are all smooth at $r=2M$. The gauge
vector is therefore nonsingular at the event horizon.  

The perturbation fields in the Regge-Wheeler gauge are then given by 
\begin{subequations} 
\begin{align} 
h_{vv}^{\rm RW} &= h_{vv}^{\rm LC} 
- 2\partial_v \xi_v^{\rm LC \to RW} 
+ \frac{2M}{r^2} \xi_v^{\rm LC \to RW} 
+ \frac{2Mf}{r^2} \xi_r^{\rm LC \to RW}, \\ 
h_{vr}^{\rm RW} &= -\partial_r \xi_v^{\rm LC \to RW} 
- \partial_v \xi_r^{\rm LC \to RW} 
- \frac{2M}{r^2} \xi_r^{\rm LC \to RW}, \\ 
h_{rr}^{\rm RW} &= -2 \partial_r \xi_r^{\rm LC \to RW}, \\ 
K^{\rm RW} &= K^{\rm LC} 
- \frac{2f}{r} \xi_r^{\rm LC \to RW} 
- \frac{2}{r} \xi_v^{\rm LC \to RW} 
+\frac{\ell(\ell+1)}{r^2} \xi_{\rm even}^{\rm LC \to RW} 
\end{align} 
\end{subequations} 
in the even-parity sector, and 
\begin{subequations} 
\begin{align} 
h_v^{\rm RW} &= h_v^{\rm LC} 
- \partial_v \xi_{\rm odd}^{\rm LC \to RW}, \\ 
h_r^{\rm RW} &= -\partial_r \xi_{\rm odd}^{\rm LC \to RW}
+ \frac{2}{r} \xi_{\rm odd}^{\rm LC \to RW} 
\end{align} 
\end{subequations} 
in the odd-parity sector. These equations show that when the 
perturbation fields are regular at $r=2M$ in the light-cone gauge,
they are also regular in the Regge-Wheeler gauge. A physically regular
perturbation is therefore nonsingular on the event horizon when it is
presented in Regge-Wheeler gauge. 

\section{From Regge-Wheeler gauge to EZ gauge} 
\label{sec:RWEZ} 

The EZ gauge conditions \cite{thompson-chen-whiting:17} are 
\begin{equation} 
j_v^{\rm EZ} = K^{\rm EZ} = G^{\rm EZ} = 0, \qquad 
h_2^{\rm EZ} = 0, 
\end{equation} 
and the last equation implies that the EZ and Regge-Wheeler gauges
coincide in the odd-parity sector. Equations (\ref{gauge_even}) reveal
that the transformation from Regge-Wheeler gauge to EZ gauge is
achieved with 
\begin{subequations}
\label{gauge_RWEZ} 
\begin{align}  
\xi^{\rm RW \to EZ}_v &= 0, \\ 
\xi^{\rm RW \to EZ}_r &= \frac{r}{2f} K^{\rm RW}, \\ 
\xi^{\rm RW \to EZ}_{\rm even} &= 0. 
\end{align} 
\end{subequations} 
The gauge vector is uniquely determined (the EZ gauge is unique), and
the factor of $f^{-1} = (1-2M/r)^{-1}$ in the radial component is the
origin of the singularity of the gauge at $r=2M$. 

The perturbation fields in the EZ gauge are given by 
\begin{subequations} 
\begin{align} 
h_{vv}^{\rm EZ} &= h_{vv}^{\rm RW} 
+ \frac{2Mf}{r^2} \xi_r^{\rm RW \to EZ}, \\ 
h_{vr}^{\rm EZ} &= h_{vr}^{\rm RW} 
- \partial_v \xi_r^{\rm RW \to EZ}
- \frac{2M}{r^2} \xi_r^{\rm RW \to EZ}, \\ 
h_{rr}^{\rm EZ} &= h_{rr}^{\rm RW} 
- 2\partial_r \xi_r^{\rm RW \to EZ}, \\
j_r^{\rm EZ} &= -\xi_r^{\rm RW \to EZ},
\end{align} 
\end{subequations} 
or 
\begin{subequations} 
\begin{align} 
h_{vv}^{\rm EZ} &= h_{vv}^{\rm RW} 
+ \frac{M}{r} K^{\rm RW}, \\ 
h_{vr}^{\rm EZ} &= h_{vr}^{\rm RW} 
- \frac{1}{f} \biggl( \frac{1}{2} r \partial_v K^{\rm RW} 
+ \frac{M}{r} K^{\rm RW} \biggr), \\ 
h_{rr}^{\rm EZ} &= h_{rr}^{\rm RW} 
- \partial_r \biggl( \frac{r}{f} K^{\rm RW} \biggr), \\ 
j_r^{\rm EZ} &= -\frac{r}{2f} K^{\rm RW}. 
\end{align} 
\end{subequations} 
The factors of $f^{-1}$ and $f^{-2}$ in the perturbation fields
(except for $h_{vv}$, which is regular) show very clearly that these
diverge when $r=2M$. A physically regular perturbation therefore
appears singular when presented in the EZ gauge. This is our main
conclusion.  

For completeness we display the transformation that takes any old
gauge to the EZ gauge. In the even-parity sector the gauge vector 
is given by 
\begin{subequations} 
\begin{align} 
\xi_v^{\rm \to EZ} &= j_v^{\rm old} 
- \frac{1}{2} r^2 \partial_v G^{\rm old}, \\ 
\xi_r^{\rm \to EZ} &= \frac{r}{2f} \biggl[ 
K^{\rm old} - \frac{2}{r} j_v^{\rm old} 
+ r \partial_v G^{\rm old} 
+ \frac{1}{2} \ell(\ell+1) G^{\rm old} \biggr], \\ 
\xi_{\rm even}^{\rm \to EZ} &= 
\frac{1}{2} r^2 G^{\rm old}, 
\end{align} 
\end{subequations} 
and it reduces to Eq.~(\ref{gauge_RWEZ}) when the old gauge is the
Regge-Wheeler gauge. The factor of $f^{-1}$ in $\xi_r^{\rm \to EZ}$
continues to imply that the EZ gauge is singular at $r=2M$. The gauge
vector is given by  
\begin{equation} 
\xi_{\rm odd}^{\rm \to EZ} = \frac{1}{2} h_2^{\rm old} 
\end{equation} 
in the odd-parity sector.  

\section{Tidally deformed black hole} 
\label{sec:tidal} 

To provide a concrete illustration of the results obtained in the
preceding sections, we consider a black hole perturbed by remote
bodies that exert tidal forces. This is a well-studied problem
\cite{poisson:04d, poisson:05, poisson-vlasov:10}, and for the most
part we shall simply import results from the literature.  

At the leading, quadrupole ($\ell=2$) order, the tidal environment is
characterized by the symmetric-tracefree Cartesian tensor 
${\cal E}_{ab}$, which is assumed to vary slowly with respect to time 
--- the tidal perturbation is formally taken to be
time-independent. The spherical-harmonic decomposition of the 
perturbation is aided by the identity 
\begin{equation} 
{\cal E}_{ab} \Omega^a \Omega^b 
= \sum_{m=-2}^2 {\cal E}_m Y^{2,m}, 
\end{equation} 
where $\Omega^a := [\sin\theta\cos\phi,
\sin\theta\sin\phi,\cos\theta]$. The identity maps the five
independent components of ${\cal E}_{ab}$ to the five harmonic
coefficients ${\cal E}_m$. The perturbation associated with ${\cal
  E}_{ab}$ is entirely in the even-parity sector.  

In the light-cone gauge, the nonvanishing perturbation fields are
given by \cite{poisson-vlasov:10} 
\begin{subequations} 
\begin{align} 
h_{vv}^{\rm LC} &= -r^2 f^2\, {\cal E}_m, \\ 
j_v^{\rm LC} &= -\frac{1}{3} r^3 f\, {\cal E}_m, \\ 
G^{\rm LC} &= -\frac{1}{3} r^2 
\biggl( 1 - \frac{2M^2}{r^2} \biggr)\, {\cal E}_m.
\end{align} 
\end{subequations} 
In the Regge-Wheeler gauge they are \cite{landry-poisson:15a} 
\begin{subequations} 
\begin{align} 
h_{vv}^{\rm RW} &= -r^2 f^2\, {\cal E}_m, \\
h_{vr}^{\rm RW} &= r^2 f\, {\cal E}_m, \\
h_{rr}^{\rm RW} &= -2 r^2\, {\cal E}_m, \\
K^{\rm RW} &= -r^2 \biggl( 1 - \frac{2M^2}{r^2} \biggr)\, {\cal E}_m. 
\end{align} 
\end{subequations} 
And in the EZ gauge they are given by 
\begin{subequations} 
\begin{align} 
h_{vv}^{\rm EZ} &= -r^2 \biggl( 1 - \frac{M}{r} \biggr) 
\biggl( 1 - \frac{2M}{r} + \frac{2M^2}{r^2} \biggr)\, {\cal E}_m, \\ 
h_{vr}^{\rm EZ} &= \frac{r^2}{f} \biggl( 1 - \frac{M}{r} \biggr) 
\biggl( 1 - \frac{2M}{r} + \frac{2M^2}{r^2} \biggr)\, {\cal E}_m, \\ 
h_{rr}^{\rm EZ} &= \frac{r^2}{f^2} \biggl( 1 - \frac{10M^2}{r^2} 
+ \frac{8M^3}{r^3} \biggr)\, {\cal E}_m, \\ 
j_r^{\rm EZ} &= \frac{r^3}{2f} 
\biggl( 1 - \frac{2M^2}{r^2} \biggr)\, {\cal E}_m. 
\end{align} 
\end{subequations} 
It is manifest that while the perturbation fields are regular at
$r=2M$ in the light-cone and Regge-Wheeler gauges, they are singular
in the EZ gauge.   

\begin{acknowledgments} 
The work reported in this article was supported by the Natural
Sciences and Engineering Research Council of Canada. 
\end{acknowledgments}

\end{document}